\documentstyle[12pt,axodraw,epsfig]{article}
\begin{document}
\newcommand{\dmchi}{\mbox{$\Delta m_{\tilde {\chi}}$}}
\newcommand{\mchi}{\mbox{$m_{\tilde {\chi}_1^0}$}}
\newcommand{\mchisq}{m_{\tilde {\chi}_1^0}^2}
\newcommand{\lsp}{\mbox{$\tilde {\chi}_1^0$}}
\newcommand{\Ochi}{\mbox{$\Omega_{\tilde \chi} h^2$}}
\newcommand{\ghcc}{\mbox{$g_{h\tilde{\chi}_1^0 \tilde{\chi}_1^0}$}}
\newcommand{\gZcc}{\mbox{$g_{Z\tilde{\chi}_1^0 \tilde{\chi}_1^0}$}}
\renewcommand{\thefootnote}{\fnsymbol{footnote}}
\pagestyle{empty}
\begin{flushright}
APCTP 97--01 \\
January 1997\\
\end{flushright}
\vspace*{1.5cm}

\begin{center}
{\Large \bf Return of the Light Higgsino\footnote{Talk given at the {\it 2nd 
International RESCEU Symposium on Dark Matter and its Direct Detection},
Tokyo, November 26--28, 1995}} \\
\vspace*{5mm}

Manuel Drees \\
{\it APCTP, Seoul National University, Seoul 151--742, Korea}
\end{center}

\begin{abstract}
It is pointed out that loop corrections involving heavy quarks and their
superpartners can re--introduce a state with 99.5\% higgsino purity
as a viable cold Dark Matter candidate. Such corrections can increase the
mass splitting between the three higgsino--like states of the MSSM by
several GeV, which results in a suppression of the co--annihilation rate
by a factor of five or more. Related corrections to the couplings of
the LSP to $Z$ and Higgs bosons can change the predicted LSP detection
rate by two orders of magnitude.
\end{abstract}
\clearpage
\setcounter{page}{1}
\pagestyle{plain}

\noindent
In models with exact R-parity, the lightest supersymmetric particle (LSP)
is stable, and is thus a particle physics candidate for the missing ``Dark
Matter'' (DM) in the Universe \cite{1}. In particular, a bino--like LSP
would have the right relic density if $m^4_{\tilde{l}_R} / \mchisq
\simeq (200 \ {\rm GeV})^2$, where $\tilde{l}_R$ stands for $SU(2)$ singlet
sleptons, which are the sfermions with the largest hypercharge; a similar
result holds for photino--like LSP.

In contrast, higgsino--like LSPs are thought to have a very small
relic density, unless their mass exceeds 0.5 TeV or so. If $\mchi >
M_W$, this is due to the very large cross sections for $\lsp \lsp
\rightarrow W^+ W^-, \ ZZ$ \cite{1}. For $\mchi < M_W$, the
annihilation cross section becomes quite small if the LSP is a nearly
pure higgsino. However, the LSP is then close in mass to the lightest
chargino and next--to--lightest neutralino. In such a situation
co--annihilation processes \cite{4} between the LSP and the only
slightly heavier higgsino--like states have to be included in the
estimate of the relic density. As pointed out in ref.\cite{5}, these
processes greatly reduce the predicted relic density of higgsino--like
LSPs, making them uninteresting as DM candidates unless the gaugino
fraction, defined as the sum of the squares of the gaugino components
of the LSP eigenvector, is at least several percent.

However, very recently it was shown \cite{6} that loop corrections \cite{7}
can change the mass splitting between the higgsino--like states of the
minimal supersymmetric standard model (MSSM) by several GeV. The authors
of ref.\cite{6} were interested in the impact of such corrections on sparticle
searches at LEP. However, since the rate for co--annihilation processes
depends {\em exponentially} on the mass differences, these corrections can
also change the prediction for the LSP relic density quite dramatically
\cite{8}. This prediction is also altered by corrections to the coupling of
the LSP to the $Z$ and, in some cases, to Higgs bosons; these couplings also
largely determine the LSP--nucleon scattering cross section, and hence the
LSP detection rate.

\begin{center}
\begin{picture}(400,180)(0,0)
\DashLine(0,90)(50,90){2}
\Line(50,90)(100,140)
\Line(50,90)(100,40)
\DashLine(80,120)(80,60){2}
\Text(15,80)[]{$Z,\Phi$}
\Text(65,115)[]{$q$}
\Text(65,65)[]{$q$}
\Text(88,90)[]{$\tilde{q}_i$}
\Text(105,150)[]{$\lsp(k_1)$}
\Text(105,30)[]{$\lsp(k_2)$}
\Text(50,10)[]{\large a)}
\DashLine(140,90)(190,90){2}
\DashLine(190,90)(220,120){2}
\DashLine(190,90)(220,60){2}
\Line(220,120)(240,140)
\Line(220,60)(240,40)
\Line(220,120)(220,60)
\Text(155,80)[]{$Z,\Phi$}
\Text(205,115)[]{$\tilde{q}_i$}
\Text(205,60)[]{$\tilde{q}_j$}
\Text(228,90)[]{$q$}
\Text(245,150)[]{$\lsp(k_1)$}
\Text(245,30)[]{$\lsp(k_2)$}
\Text(190,10)[]{\large b)}
\DashLine(280,90)(330,90){2}
\Line(330,90)(380,140)
\Line(330,90)(380,40)
\DashCArc(355,115)(12,45,225){2}
\Text(295,80)[]{$Z$}
\Text(359,109)[]{$q$}
\Text(355,136)[]{$\tilde{q}_i$}
\Text(385,150)[]{$\lsp(k_1)$}
\Text(385,30)[]{$\lsp(k_2)$}
\Text(333,106)[]{$\tilde{\chi}_2^0$}
\Text(330,10)[]{\large c)}
\end{picture}
\end{center}

\noindent
{\bf Fig.~1:} {\small Quark--squark loop corrections to the coupling of a pair
of LSPs to a $Z$ or Higgs boson. The LSP momenta $k_1$ and $k_2$ point
towards the vertex. Note that both senses of the ``Dirac arrow'' (flow of
fermion number) have to be added, since the LSP is a Majorana fermion.
There is also a diagram of type c) with a quark--squark bubble on the
other neutralino line. There are two squark mass eigenstate with
a given flavor.}\\
\vspace*{5mm}

We therefore computed \cite{8} the vertex corrections of Figs.~1 in
addition to the corrections to the masses given in
refs.\cite{7,6}. Note that the off--diagonal wave function
renormalization diagram of Fig.~1c, which only contributes to the $Z
\lsp \lsp$ coupling, is closely related to the corrections to the
neutralino mass matrix; we found that this contribution usually
dominates the correction to this coupling. In contrast, the
potentially largest contribution to the couplings of the scalar Higgs
bosons to the LSP comes from Fig.~1b; this contribution depends
sensitively on the soft SUSY breaking $A$ parameters that also appear
in the off--diagonal entries of the squark mass matrices
\cite{1}. This is also true for the corrections to the mass
splittings, which vanish if the two squarks of a given flavour are
mass--dengenerate or if the $\tilde{q}_L - \tilde{q}_R$ mixing angle
goes to zero.

This is demonstrated in Fig.~2, which shows the chargino--LSP mass splitting
(solid), the axial--vector $Z \lsp \lsp$ coupling (long dashed), and the
coupling of the LSP to the light scalar Higgs boson $h^0$ (short dashed);
all results have been normalized such that they can be plotted to a common
scale. In all three cases the tree--level prediction is very close to the
one--loop corrected estimate for $A=0$. We see that the mass splitting can
change by about $\pm 4$ GeV, while the $Z \lsp \lsp$ coupling changes by
about a factor of three as $A$ is varied across its allowed range. Since for
the given choice of parameters (in particular, $\mu < 0$) the tree--level
prediction for \ghcc\ is very small, the loop corrections can even flip the
sign of this coupling.

\setcounter{figure}{1}
\begin{figure}
\vspace*{2.6cm}
\centerline{\epsfig{file=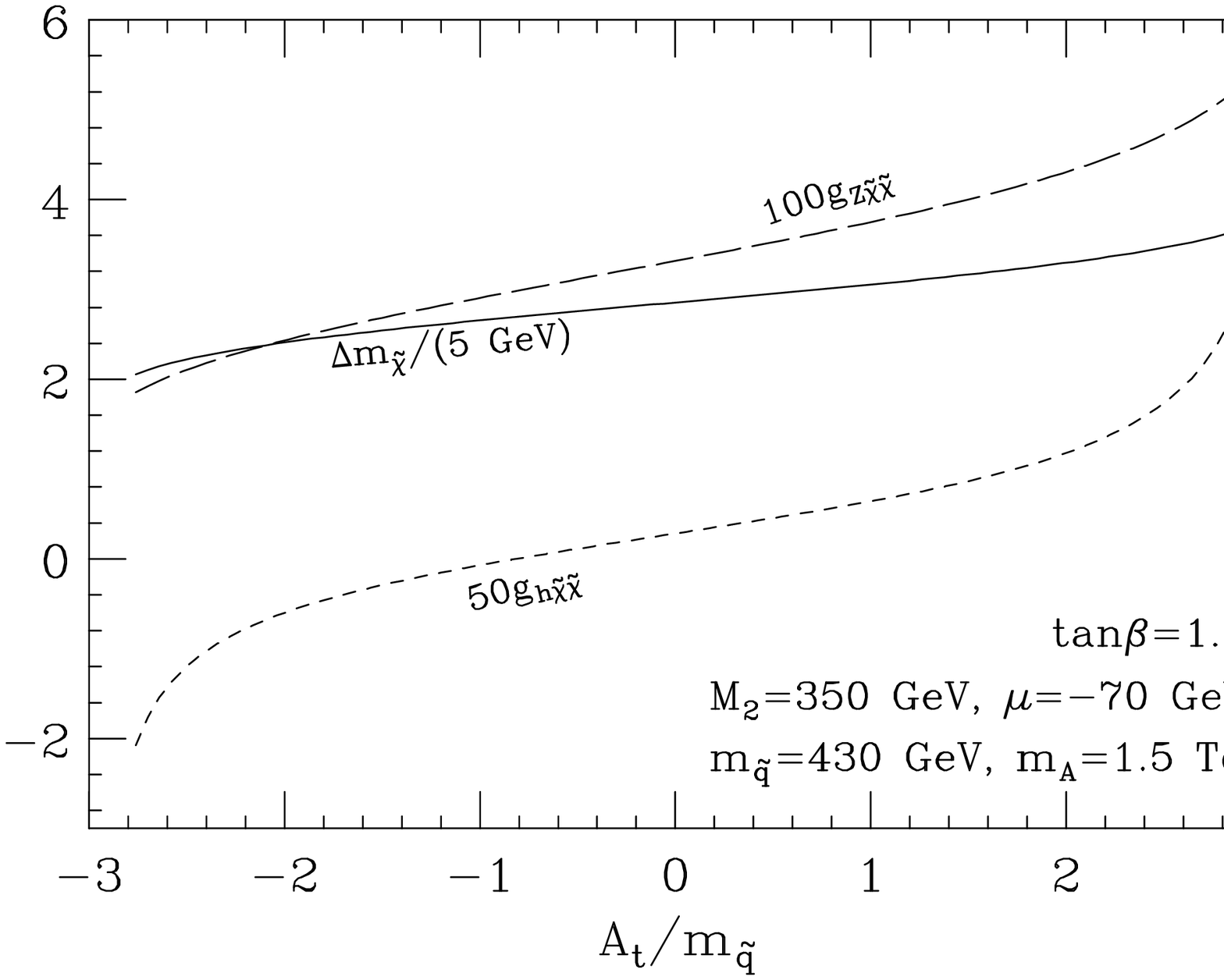,height=7.7cm}}
\caption
{The chargino--LSP mass difference (solid), the 
axial--vector $Z \lsp \lsp$ coupling (long dashed), and the $h^0 \lsp \lsp$
coupling (short dashed) are shown as a function of the soft breaking $A$
parameter, including one--loop corrections involving Yukawa couplings.}
\end{figure}

Fig.~3 shows predictions for the LSP relic density as a function of the
gaugino fraction. Since the LSP mass has been kept fixed at 70 GeV, both the
mass $M_2$ of the $SU(2)$ gauginos\footnote{We assume gaugino mass unification}
and the higgsino mass parameter $\mu$ vary along the curves. The dotted curve
has been obtained ignoring loop corrections to the higgsino masses and
couplings; the other two curves include these corrections, with $A$ being
close to its upper and lower limit, respectively. We see that in the region of
high higgsino purity, the corrections can either increase or decrease the
predicted LSP relic density by more than a factor of five. If $A$ is large
and positive, a state with 99.9\% higgsino purity can form galactic DM haloes
($\Ochi \geq 0.02$), while a state with 99.5\% higgsino purity can form all
cold DM in the recently popular mixed DM models ($\Ochi \geq 0.15$).

\begin{figure}
\vspace*{2.6cm}
\centerline{\epsfig{file=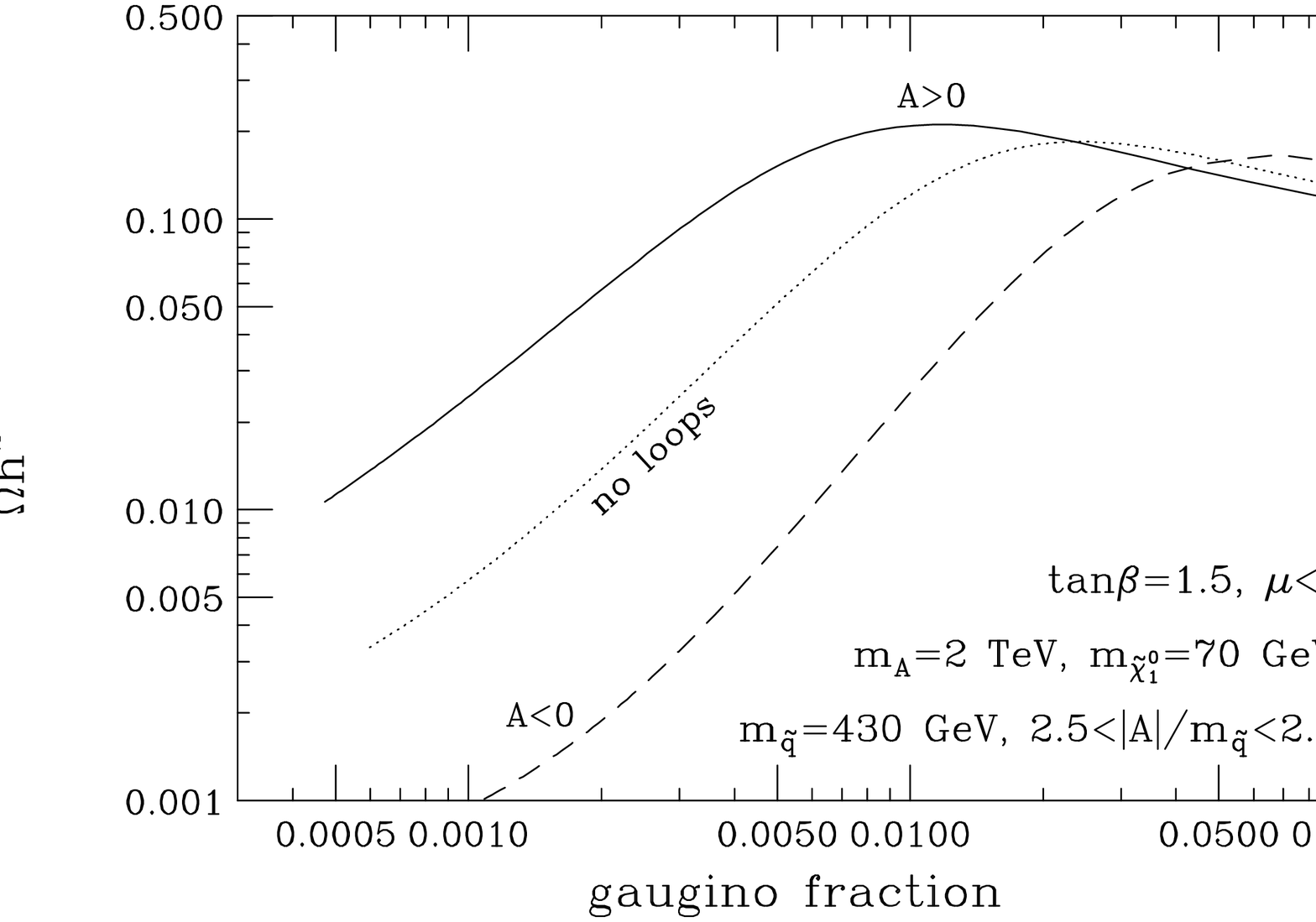,height=7.7cm}}
\caption{The LSP
relic density \Ochi\ is shown as a function of the gaugino fraction.
These results are for a fixed LSP mass, so that both
$M_2$ and $\mu$ vary along the text. Further, $|A|$ has been
decreased from $2.7 m_{\tilde q}$ to $2.5 m_{\tilde q}$ as $M_2$ was
increased from about 150 GeV to 1 TeV.}
\end{figure}

Finally, Fig.~4 shows the dependence of the predicted LSP detection rate in an
isotopically pure ${}^{76}$Ge detector on the $A$ parameter. The dotted curve
again holds in the absence of the loop corrections of Figs.~1. There is still
some dependence on $A$, due to top--stop loop corrections to $m_{h^0}$
\cite{10}. Clearly the $A-$dependence becomes much stronger once the
corrections of Figs.~1 are included (solid curve). If $|A|$ is near its
upper limit, these corrections increase the predicted counting rate by about
two orders of magnitude. However, they can also lead to an exactly vanishing
cross section for LSP scattering off spinless nuclei. This happens near the
point where $\ghcc=0$ (see Fig.~2).\footnote{There are also small $q - 
\tilde{q}$ loop contributions \cite{1} to the scattering matrix element;
therefore the complete matrix element vanishes at a small positive value
of \ghcc, rather than at $\ghcc=0$.}

In summary, loop corrections re--introduce a 99.5\% pure higgsino state as
viable CDM candidate. These corrections can also increase the expected LSP
detection rate by two orders of magnitude, if $\mu < 0$. In both cases the
sign of the corrections can also be opposite, however, suppressing the
relic density, and perhaps even reducing the LSP scattering cross section
off spinless nuclei to zero.
\clearpage
\noindent
{\bf \large Acknowledgements}
I thank my collaborators Mihoko Nojiri, D.P. Roy and Youichi Yamada for an
enjoyable collaboration. I also thank the members of the KEK theory division
for their hospitality while this report was written.

\begin{figure}
\vspace*{2.5cm}
\centerline{\epsfig{file=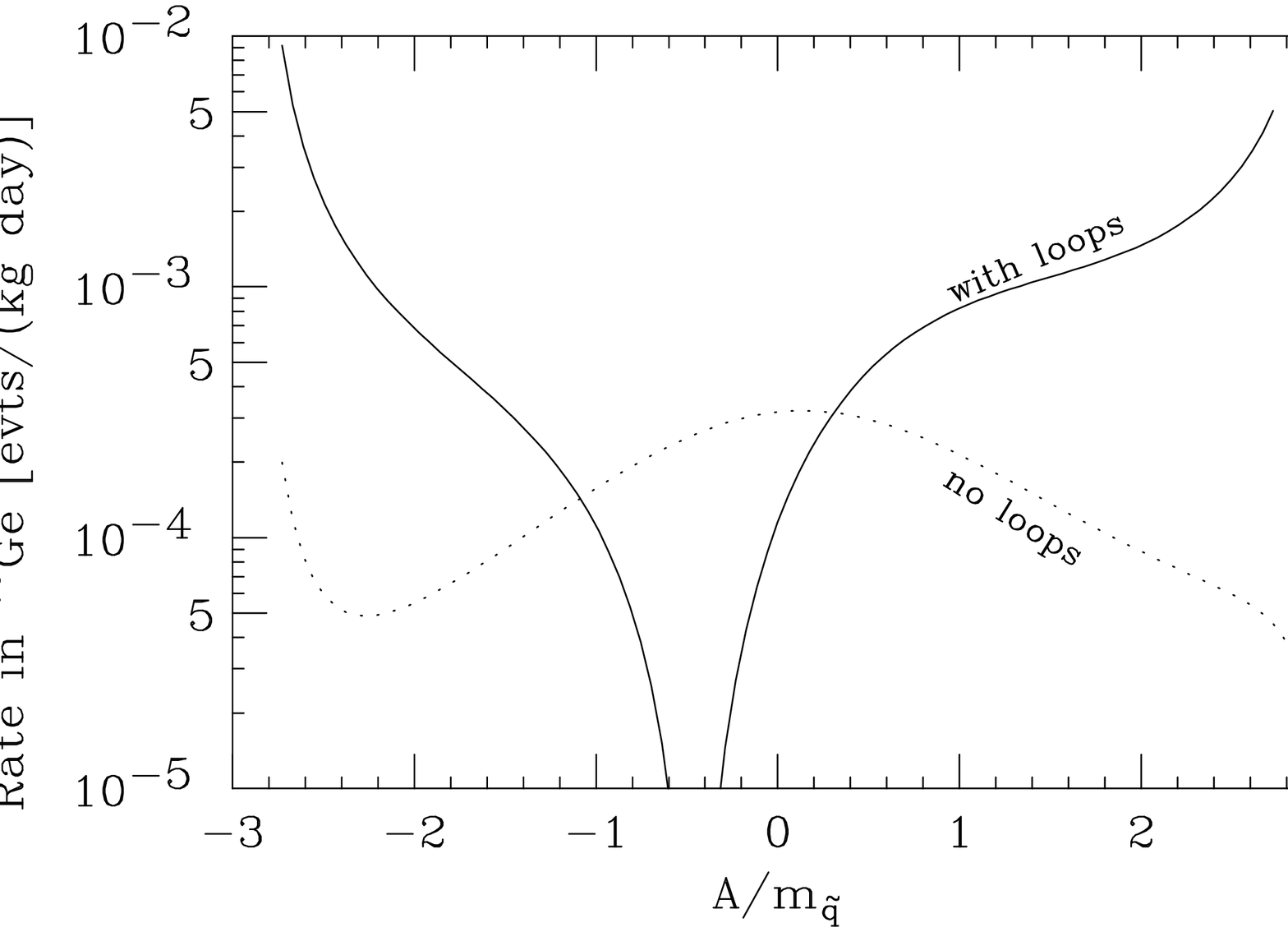,height=7.cm}}
\caption{The expected LSP detection rate in a ${}^{76}$Ge
detector is plotted as a function of $A$, with (solid) and without
(dotted) the corrections depicted in Figs.~1. The values of the other
parameters are: $m_{\tilde{\chi}^0_1}=70 \ {\rm GeV}, 
\ m_{\tilde q}=430 \ {\rm GeV}, \ M_2 = 300 \ {\rm GeV}, \ m_A = 1.5 \ 
{\rm TeV}$, and $\tan \! \beta=1.5$.}
\end{figure}

 \end{document}